\documentclass[]{aa}
\usepackage{graphicx}
\usepackage{txfonts}
\begin{document}
\title{Detection of circumstellar disks around nearby young brown dwarf candidates}

   \author{N. Phan-Bao
       \inst{1,2}
       \and
       M.S. Bessell
       \inst{3}
       \and
       E.L. Mart\'{\i}n
       \inst{4}    
       }

\offprints{N.~Phan-Bao}

\institute{Department of Physics, International University, Ho Chi Minh City, Vietnam. \\
           \email{pbngoc@hcmiu.edu.vn}
           \and
           Vietnam National University, Ho Chi Minh City, Vietnam.
           \and 
           Research School of Astronomy and Astrophysics, Australian National University, 
           Cotter Rd, Weston, ACT 2611, Australia.
          \and
           Instituto de Astrof\'{\i}sica de Canarias, Spain.
    	  }

      \date{Received; accepted}

\abstract 
    {It is important to detect and study circumstellar disks around late-M and brown dwarfs of nearby young associations to understand how these very low-mass objects form
     and how rocky planets form around them.
     The detection of new very low-mass members of nearby young associations will also significantly
     improve our current understanding of young associations. 
     }
    {We searched for new young very low-mass members with circumstellar disks
     in a sample of 3928 candidates.}
    {We constructed spectral energy distributions of all candidates using observational
     photometric data from DENIS, 2MASS, and WISE and trigonometric parallaxes from Gaia
     to detect infrared excess emission that indicates the presence of circumstellar disks. 
     We then followed up spectroscopic observations of candidates to search for lithium
     to confirm their youth. 
     The H$\alpha$ emission line was used to detect accretion.
    }
    {We detected 23 among the 3928 candidates with circumstellar disks:
      Ten objects are new, and 13 were previously reported in the literature.
      Our mass estimates also indicate that 21 are brown dwarf candidates
     and 2 are very low-mass stars.
     DENIS~J0534552$-$104808 has a Gaia distance of
     238~pc and might be the first brown dwarf
     candidate member of a foreground population in front of the Orion D cloud. 
     This foreground population is probably associated with the supergiant $\kappa$ Ori. 
     Based on our spectroscopic observations, we detected lithium in 11 candidates. 
     We also identified seven accretors and one potential accretor. 
     The intense long-lived accretion detected in DENIS-P~J0500245$-$333042, a 20 Myr old
     brown dwarf candidate may be additional evidence to favor
     the formation of rocky planets around very low-mass objects.
     }   
{}

\authorrunning{Phan-Bao et al.}
\titlerunning{Circumstellar dust around young brown dwarf candidates}

  \maketitle

\section{Introduction}

Late-M and brown dwarf (BD) populations in nearby young associations
at different ages offer us excellent opportunities to study the formation mechanism
of these very low-mass (VLM) objects and the planet formation around them.
The search for VLM BDs close to the BD-planet boundary has also led to the discovery of
planetary mass objects
\citep{zapatero,martin25} that challenge our understanding of the formation mechanism of
free-floating planets.

The Gaia DR3 archive \citep{gaia23} contains about 1.8 billion sources
with high-precision proper motions and trigonometric parallaxes.
With a limiting magnitude of about 20.7 in $G$ band, Gaia can detect
an M5.5 dwarf with a $G$-band absolute magnitude
of $M_{\rm G} = 12.25$ \citep{cif} at $\sim$490~pc and an M8.0 with $M_{\rm G}= 15.19$
at $\sim$127~pc. 
Proper motions and parallaxes are a key piece in identifying
new members of nearby young associations. Therefore, Gaia data
are an invaluable resource to study VLM populations in nearby young associations
(e.g., \citealt{gagne18b,sarro}).

Based on photometric and/or spectroscopic distances,
thousands of late-M dwarfs have been identified in the past decades
as nearby ($\leq$ 30 pc) VLM stars in the field
(e.g., \citealt{pb01,pb03,pb08a,cruz02,reid03,martin10}).
Many of them are in fact young VLM stars and BDs of nearby young associations, however.
The misidentification of these late-M dwarfs is due to the lack
of constraints on the age, such as the presence of lithium and trigonometric parallaxes.
In our previous search for young BD candidates in a sample of
85 nearby late-M dwarfs (M5.0$-$M9.0), which were originally identified over
5700 square degrees in the Deep Near Infrared Survey of
the Southern Sky (DENIS) database with photometric distances $\leq$ 30~pc,
we discovered four new young BD candidates and one new young VLM star
in Tucana-Horologium, Columba, $\beta$ Pic, and Upper Scorpius \citep{pb17,dat}.
One of them, a young BD (DENIS-P~J1538317$-$103850), shows sporadic and intense accretion.
The detections imply that nearby field late-M dwarfs identified
in the DENIS database may in fact be young BD members of nearby young associations. 
Following the previous detections, we expanded our search to the entire
DENIS database over 13500 square degrees at high Galactic latitude to identify new
young BDs with circumstellar disks. 

We present our sample in Sec.~2. The detection of new BD candidates and VLM stars in
nearby young associations and spectroscopic follow-up is presented in Sec.~3.
We summarize our results in Sec.~4.

\section{Sample}

In our previous paper \citep{dat}, we presented our search for
circumstellar disks around 85 nearby late-M dwarfs.
These nearby late-M dwarfs were previously identified 
over 5700 square degrees in the DENIS database
at high Galactic latitude with $|b| \geq 30^{\circ}$ and photometric distances
$d_{\rm phot} \leq 30$~pc.
The photometric distances were estimated
using the $I-J$ versus $M_{I}$ relation for
old M dwarfs with known trigonometric parallaxes
as given in \citet{pb03}. 

In this paper, we expanded our search to the entire DENIS database of
over about 13500 square degrees,
down to lower Galactic latitude with $|b| \geq 15^{\circ}$.
For this search, we focused on 
distant late-M dwarf candidates with $d_{\rm phot} > 30$~pc.
The photometric distances were estimated using the same
$I-J$ versus $M_{I}$ relation as for the previous search.
Our sample consisted of 3928 late-M dwarf candidates
with a DENIS color in the range
of $2.0 \leq I-J < 3.0$ ($\sim$M5$-$M8.5; \citealt{pb06a}).
We then cross-identified these candidates in the 2MASS \citep{2mass}
and AllWISE \citep{cutri14} catalogs.
The cross-identification resulted in 3515 candidates for which mid-infrared photometric data
are available in WISE.
 
We followed the same method as in the previous search for infrared (IR) excesses
in late-M dwarfs (see also \citealt{dat}).
With photometric data from 
DENIS $I$-band magnitude, 2MASS $J$, $H$, and $K_{\rm S}$-band magnitudes,
and WISE $W1$, $W2$, $W3$, and $W4$-band magnitudes,
we constructed
the spectral energy distribution (SED) of all 3515 candidates.
We converted magnitudes into fluxes using 
zero-magnitude flux values for DENIS \citep{fouque}, 2MASS \citep{cohen}, and WISE \citep{wright}.
We then used the BT-Settl model atmospheres with solar metallicity and surface gravities 
$\log g$ from 3.0 to 5.5 \citep{allard13}
to determine the best fits of the models to the observational data.
Distances derived from trigonometric parallaxes were used if available in Gaia \citep{gaia16,gaia23}.
We selected any candidates with observed WISE IR fluxes higher than 3$\sigma$ above the BT-Settl model fluxes,
where $\sigma$ is the error on the observed fluxes. In total, 160 candidates
matched our selection criterion. 
We then further examined the candidates visually in
all four WISE band images\footnote{https://irsa.ipac.caltech.edu/applications/wise/}.
Thirty-seven duplicate sources were removed.
The $W3$ and $W4$-band fluxes of a source
are reliable if the source is clearly visible in the images of these bands.
After our visual inspection of the WISE images of all 160 candidates, 
we finally had 23 candidates with a reliable detection of an IR excess.
We also examined the possibility that $W3$ and $W4$-band fluxes
are contaminated by background objects using SERC\footnote{https://archive.stsci.edu/cgi-bin/dss\_plate\_finder}
and 2MASS\footnote{https://irsa.ipac.caltech.edu/applications/2MASS/IM/interactive.html} images.
No contaminant was found within the largest aperture of 16.$''$5 ($W4$) of the four WISE bands.

\begin{table*}
  {\small
   \caption{DENIS and 2MASS photometric data of 23 M dwarfs with IR excess.}
    \label{nearir}
  $$
   \begin{tabular}{lllllll}
   \hline 
   \hline
   \noalign{\smallskip}
  DENIS-P name & $\alpha_{\rm 2000}$ & $\delta_{\rm 2000}$ & $I_{\rm DENIS}$ & $J_{\rm 2MASS}$ &  $H_{\rm 2MASS}$ & $K_{\rm 2MASS}$   \\
               &                   &                    &(mag)          & (mag)        &   (mag)         &  (mag)           \\
         \noalign{\smallskip}
\hline
J0322152$-$354719 & 03 22 15.28 & $-$35 47 19.7 & 15.11$\pm$0.05 & 12.965$\pm$0.027 & 12.302$\pm$0.028 & 12.024$\pm$0.027 \\
J0338011$-$333421 & 03 38 01.13 & $-$33 34 21.4 & 15.72$\pm$0.05 & 13.548$\pm$0.027 & 12.973$\pm$0.030 & 12.612$\pm$0.034 \\
J0338520$-$264615 & 03 38 52.01 & $-$26 46 15.3 & 15.53$\pm$0.05 & 13.591$\pm$0.027 & 13.046$\pm$0.026 & 12.753$\pm$0.027 \\
J0500245$-$333042 & 05 00 24.58 & $-$33 30 42.0 & 16.61$\pm$0.08 & 14.156$\pm$0.026 & 13.568$\pm$0.022 & 13.190$\pm$0.030 \\
J0516542$-$022433 & 05 16 54.22 & $-$02 24 33.2 & 16.11$\pm$0.06 & 14.035$\pm$0.027 & 13.400$\pm$0.041 & 12.992$\pm$0.035 \\
J0521590$-$080238 & 05 21 59.07 & $-$08 02 38.2 & 16.02$\pm$0.09 & 13.945$\pm$0.026 & 13.323$\pm$0.027 & 12.982$\pm$0.033 \\
J0524010$-$043715 & 05 24 01.00 & $-$04 37 15.2 & 15.26$\pm$0.04 & 13.353$\pm$0.023 & 12.767$\pm$0.026 & 12.421$\pm$0.021 \\
J0534552$-$104808 & 05 34 55.27 & $-$10 48 08.4 & 15.43$\pm$0.06 & 13.159$\pm$0.023 & 12.537$\pm$0.034 & 12.125$\pm$0.020 \\
J0844091$-$783346 & 08 44 09.20 & $-$78 33 46.0 & 14.83$\pm$0.04 & 12.505$\pm$0.024 & 11.976$\pm$0.022 & 11.618$\pm$0.024 \\
J0856138$-$134224 & 08 56 13.84 & $-$13 42 24.1 & 16.48$\pm$0.12 & 13.602$\pm$0.026 & 12.976$\pm$0.029 & 12.489$\pm$0.024 \\
J1334530$-$461920 & 13 34 53.05 & $-$46 19 20.3 & 16.53$\pm$0.07 & 13.976$\pm$0.029 & 13.385$\pm$0.032 & 13.000$\pm$0.034 \\
J1342325$-$441356 & 13 42 32.60 & $-$44 13 56.3 & 15.56$\pm$0.04 & 13.403$\pm$0.026 & 12.792$\pm$0.025 & 12.450$\pm$0.028 \\
J1344474$-$433658 & 13 44 47.47 & $-$43 36 58.5 & 15.59$\pm$0.06 & 13.543$\pm$0.033 & 12.976$\pm$0.029 & 12.641$\pm$0.030 \\
J1403297$-$335456 & 14 03 29.70 & $-$33 54 56.9 & 14.91$\pm$0.03 & 12.798$\pm$0.021 & 12.267$\pm$0.024 & 11.982$\pm$0.025 \\
J1406468$-$415104 & 14 06 46.85 & $-$41 51 04.7 & 16.19$\pm$0.07 & 13.615$\pm$0.024 & 12.980$\pm$0.022 & 12.631$\pm$0.027 \\
J1448016$-$385926 & 14 48 01.61 & $-$38 59 26.4 & 15.47$\pm$0.14 & 12.757$\pm$0.023 & 12.144$\pm$0.022 & 11.817$\pm$0.024 \\
J1506573$-$341438 & 15 06 57.38 & $-$34 14 38.0 & 15.64$\pm$0.05 & 13.505$\pm$0.023 & 12.734$\pm$0.025 & 12.163$\pm$0.021 \\
J1546269$-$244322 & 15 46 26.97 & $-$24 43 22.6 & 14.81$\pm$0.06 & 12.644$\pm$0.023 & 12.014$\pm$0.023 & 11.702$\pm$0.023 \\
J1546543$-$255652 & 15 46 54.33 & $-$25 56 52.3 & 15.21$\pm$0.06 & 12.851$\pm$0.026 & 12.133$\pm$0.026 & 11.790$\pm$0.023 \\
J1558546$-$181324 & 15 58 54.66 & $-$18 13 24.1 & 15.35$\pm$0.05 & 13.216$\pm$0.028 & 12.557$\pm$0.028 & 12.207$\pm$0.031 \\
J1605540$-$181844 & 16 05 54.07 & $-$18 18 44.1 & 15.96$\pm$0.07 & 13.422$\pm$0.028 & 12.840$\pm$0.027 & 12.507$\pm$0.030 \\
J1853073$-$360516 & 18 53 07.36 & $-$36 05 16.3 & 14.72$\pm$0.04 & 12.702$\pm$0.027 & 12.191$\pm$0.022 & 11.871$\pm$0.021 \\
J1856402$-$365520 & 18 56 40.25 & $-$36 55 20.4 & 15.82$\pm$0.05 & 13.670$\pm$0.026 & 13.077$\pm$0.031 & 12.818$\pm$0.033 \\
      \noalign{\smallskip}
\hline
    \noalign{\smallskip}
    \hline 
   \end{tabular}
  $$
  }
\end{table*}

\begin{table*}
  {\small
   \caption{WISE photometric and Gaia data of 23 M dwarfs.}
    \label{ir}
  $$
   \begin{tabular}{lllllllll}
   \hline 
   \hline
   \noalign{\smallskip}
  D* name& $W$1               & $W$2               & $W$3               & $W$4                &  $d$   & $\pi$     & $\mu$RA      & $\mu$DE     \\
            & (mag)            & (mag)            & (mag)            &  (mag)         &  (pc)& (mas)   & (mas/yr) & (mas/yr) \\
         \noalign{\smallskip}
\hline
0322$-$3547 & 11.757$\pm$0.022 & 11.504$\pm$0.021 & ~~9.951$\pm$0.035 &  8.248$\pm$0.182 & 104.4 & ~~9.5772$\pm$0.1426 &\,~~37.492$\pm$0.119 &~~$-$3.249$\pm$0.162 \\
0338$-$3334 & 12.272$\pm$0.025 & 12.053$\pm$0.024 & 11.004$\pm$0.087  &  8.938$\pm$0.328 & 102.9 & ~~9.7178$\pm$0.1099 &\,~~35.373$\pm$0.096 &~~$-$4.571$\pm$0.116 \\
0338$-$2646 & 12.531$\pm$0.024 & 12.215$\pm$0.022 & ~~9.894$\pm$0.043 &  7.576$\pm$0.124 & 116.7 & ~~8.5682$\pm$0.0708 &\,~~31.871$\pm$0.041 &~~$-$6.906$\pm$0.065 \\
0500$-$3330 & 12.945$\pm$0.023 & 12.690$\pm$0.024 & 11.181$\pm$0.100  &  8.573$\pm$0.219 & ~~96.8& 10.3273$\pm$0.1192  &\,~~27.696$\pm$0.120 &\,~~~~3.440$\pm$0.139 \\
0516$-$0224 & 12.737$\pm$0.024 & 12.374$\pm$0.024 & 10.452$\pm$0.074  &  8.280$\pm$0.283 &       &                     &\,~~~~9$\pm$4$^{\rm (a)}$&\,~~~~2$\pm$13$^{\rm (a)}$ \\
0521$-$0802 & 12.733$\pm$0.023 & 12.248$\pm$0.023 & 10.127$\pm$0.064  &  8.230$\pm$0.246 & 365.3 & ~~2.7377$\pm$0.1135 &~~$-$0.014$\pm$0.095 &~~$-$3.803$\pm$0.083 \\
0524$-$0437 & 12.289$\pm$0.023 & 12.004$\pm$0.023 & 10.443$\pm$0.071  &  8.312$\pm$0.387 &       &                     &\,~~~~8$\pm$15$^{\rm (a)}$&\,~~10$\pm$3$^{\rm (a)}$ \\
0534$-$1048 & 11.461$\pm$0.026 & 10.740$\pm$0.026 & ~~8.631$\pm$0.028 &  6.511$\pm$0.067 & 238.1 & ~~4.1994$\pm$0.3128 &\,~~~~2.511$\pm$0.289&~~$-$4.320$\pm$0.264 \\
0844$-$7833 & 11.214$\pm$0.023 & 10.701$\pm$0.020 & ~~9.041$\pm$0.026 &  7.212$\pm$0.072 & ~~99.6& 10.0441$\pm$0.0393  & $-$29.844$\pm$0.054 & \,~~26.262$\pm$0.049 \\
0856$-$1342 & 12.168$\pm$0.023 & 11.639$\pm$0.022 & ~~9.823$\pm$0.047 &  8.326$\pm$0.295 & ~~53.8& 18.5877$\pm$0.1536  & $-$58.872$\pm$0.145 & $-$22.400$\pm$0.105 \\
1334$-$4619 & 12.272$\pm$0.023 & 11.620$\pm$0.022 & 10.084$\pm$0.047  &  8.397$\pm$0.234 & 129.8 & ~~7.7054$\pm$0.1384 & $-$26.729$\pm$0.144 & $-$19.262$\pm$0.115 \\
1342$-$4413 & 12.105$\pm$0.023 & 11.668$\pm$0.022 & 10.191$\pm$0.049  &  8.761$\pm$0.306 & 128.2 & ~~7.7993$\pm$0.1108 & $-$27.209$\pm$0.112 & $-$20.257$\pm$0.091 \\
1344$-$4336 & 12.281$\pm$0.022 & 11.698$\pm$0.021 & ~~9.723$\pm$0.033 &  7.780$\pm$0.144 & 141.1 & ~~7.0889$\pm$0.1297 & $-$26.062$\pm$0.082 & $-$19.841$\pm$0.083 \\
1403$-$3354 & 11.821$\pm$0.022 & 11.526$\pm$0.021 & ~~9.573$\pm$0.037 &  7.548$\pm$0.122 & 121.0 & ~~8.2654$\pm$0.0630 & $-$26.639$\pm$0.073 & $-$22.261$\pm$0.088 \\
1406$-$4151 & 12.291$\pm$0.023 & 11.822$\pm$0.022 & 10.604$\pm$0.067  &  8.646$\pm$0.270 &       &                     &\,~~~~9$\pm$36$^{\rm (a)}$& $-$53$\pm$33$^{\rm (a)}$ \\
1448$-$3859 & 10.936$\pm$0.022 & 10.359$\pm$0.021 & ~~9.160$\pm$0.030 &  7.604$\pm$0.157 & 119.0 & ~~8.4019$\pm$0.0675 & $-$28.156$\pm$0.070 & $-$26.504$\pm$0.057 \\
1506$-$3414 & 11.664$\pm$0.023 & 10.995$\pm$0.021 & ~~9.461$\pm$0.039 &  7.473$\pm$0.147 & 124.8 & ~~8.0117$\pm$0.0985 & $-$24.657$\pm$0.103 & $-$26.106$\pm$0.089 \\
1546$-$2443 & 11.520$\pm$0.022 & 11.256$\pm$0.021 & ~~9.747$\pm$0.053 &  7.041$\pm$0.102 & 147.6 & ~~6.7772$\pm$0.0570 & $-$15.609$\pm$0.066 & $-$24.266$\pm$0.044 \\
1546$-$2556 & 11.356$\pm$0.055 & 10.864$\pm$0.022 & ~~9.060$\pm$0.061 &  6.480$\pm$0.093 & 150.1 & ~~6.6615$\pm$0.0699 & $-$13.745$\pm$0.088 & $-$22.947$\pm$0.060 \\
1558$-$1813 & 11.955$\pm$0.023 & 11.602$\pm$0.022 & 10.091$\pm$0.067  &  8.095$\pm$0.262 & 149.9 & ~~6.6728$\pm$0.0992 & $-$10.300$\pm$0.109 & $-$21.216$\pm$0.088 \\
1605$-$1818 & 12.125$\pm$0.023 & 11.694$\pm$0.023 & ~~9.630$\pm$0.045 &  7.560$\pm$0.159 & 146.0 & ~~6.8485$\pm$0.1323 &~~$-$8.993$\pm$0.183 & $-$20.896$\pm$0.095 \\
1853$-$3605 & 11.683$\pm$0.023 & 11.394$\pm$0.021 & ~~9.439$\pm$0.043 &  7.599$\pm$0.170 & 159.3 & ~~6.2771$\pm$0.0561 &\,~~~~2.068$\pm$0.062& $-$25.081$\pm$0.049 \\
1856$-$3655 & 12.480$\pm$0.025 & 12.040$\pm$0.023 & 10.309$\pm$0.088  &  8.079$\pm$0.260 & 146.8 & ~~6.8098$\pm$0.0976 &\,~~~~3.761$\pm$0.104& $-$28.070$\pm$0.094 \\
      \noalign{\smallskip}
\hline
    \noalign{\smallskip}
    \hline 
   \end{tabular}
  $$
  \begin{list}{}{}
  \item[] 
    {\bf Notes.} \\
    D*: DENIS; the distances $d$ were derived from Gaia trigonometric parallaxes $\pi$;  \\
    $^{\rm (a)}$ our proper motion measurements if not available in Gaia.\\ 
  \end{list}
  }
\end{table*}

\begin{figure}
   \centering
    \includegraphics[width=8cm,angle=-90]{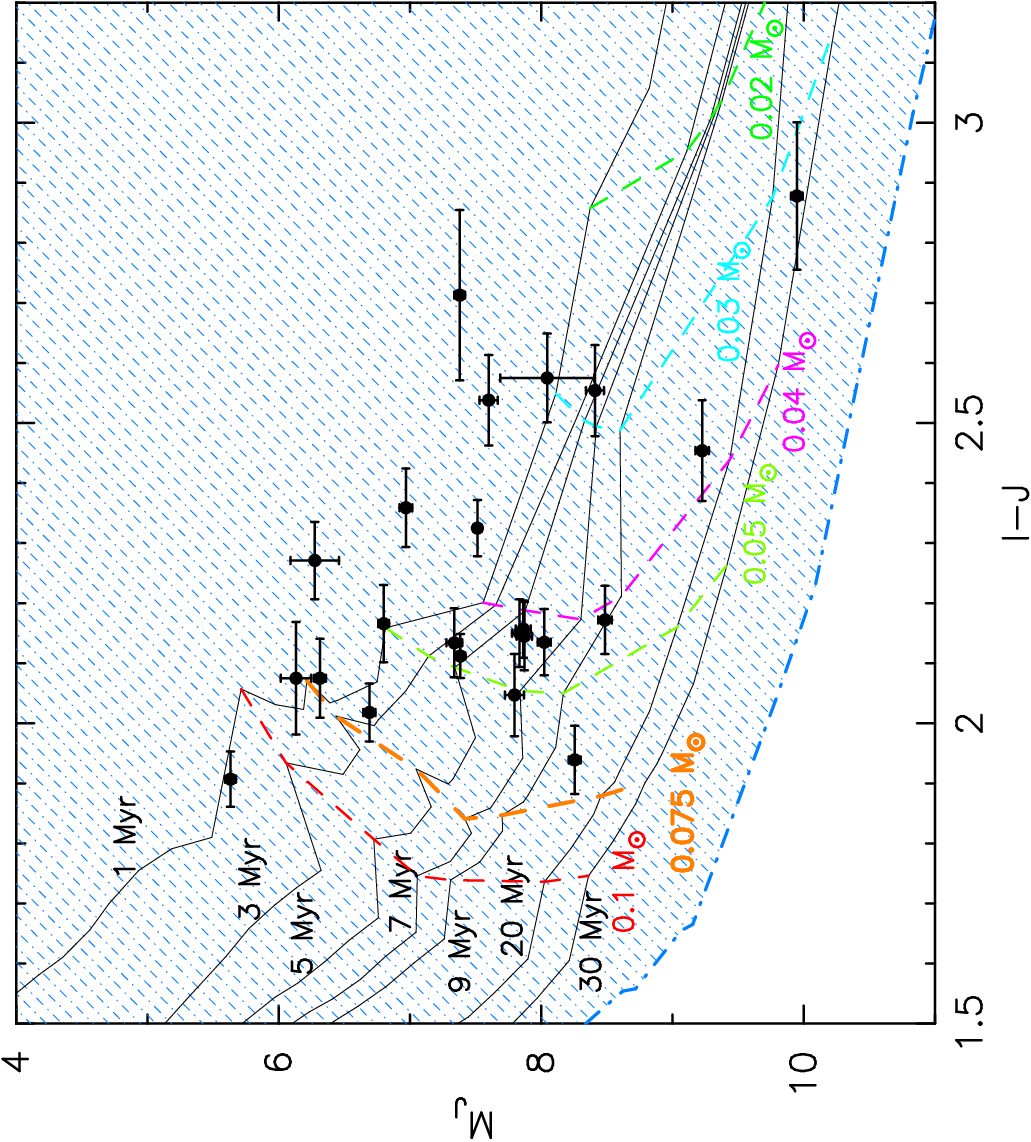}
    \caption{2MASS $J$-band absolute magnitude vs. $I-J$ color diagram for 23 candidates
      with available Gaia parallaxes. The DENIS $I$-band and 2MASS $J$-band photometric data are
      listed in Table~\ref{nearir}. 
      Isochrones and mass tracks from the BT-Settl CIFIST 2011-2015 models \citep{baraffe15}
      are plotted.
      The hatched blue area indicates the region
      in which lithium is not completely depleted and can  be detected in the stellar atmosphere.
}
\label{f1}
\end{figure}

\section{Results and discussion}

Table~\ref{nearir} and Table~\ref{ir} list 
the DENIS, 2MASS, and WISE photometric data and the Gaia
trigonometric parallaxes and proper motions
of these 23 candidates.
Three candidates lack Gaia proper motions:  DENIS0516$-$0224, DENIS0524$-$0437, and DENIS1406$-$4151.
We then measured the proper motion of these three candidates
based on the plates from the Digital Sky Survey (DSS) server,
DENIS and 2MASS positions as described in \citet{pb01}.
The time baselines span from 6 to 48 yr. Our measurements are included in Table~\ref{ir}.
Using these available data and the BT-Settl CIFIST 2011-2015 models \citep{baraffe15},
we then determined the mass and age of the candidates and their membership in
young associations. 

We also followed-up low-resolution spectroscopic observations of 11 candidates
to determine their spectral types, and we searched for the Li~I absorption line at 6708~\AA~
to confirm their youth.
Six of these 11 candidates were spectroscopically observed with medium resolution
to search for the presence of accretion. 
We present our results in the following sections. 

\subsection{Determination of age, mass, and membership in young associations}
To determine the age and mass of a candidate, we used the Gaia trigonometric
distance and 2MASS $J$-band magnitude to derive the $J$-band absolute magnitude of the candidate.
Using the 2MASS $J$-band absolute magnitude and the $I_{\rm DENIS}-J_{\rm 2MASS}$ (hereafter $I-J$) color,
we estimated the age from the closest isochrone of the BT-Settl CIFIST 2011-2015 models
to the location of the candidate in the ($M_{\rm J}$, $I-J$) diagram.
Figure~\ref{f1} shows the $J$-band absolute magnitude versus $I-J$ color diagram for
all 23 candidates with isochrones and mass tracks of the models.
With the models for the estimated age,
we then used the $J$-band absolute magnitude and its error
to determine the mass of the candidate and the uncertainty on the estimated mass, respectively.
Table~\ref{mass} shows our estimates of age, mass, and mass uncertainty.
When the error of the $I-J$ color is taken into account,
the mass uncertainty is significantly larger than the listed value. 
For DENIS0516$-$0224, DENIS0524$-$0437, and DENIS1406$-$4151,
no Gaia trigonometric parallaxes are available.
We assumed that DENIS0516$-$0224 and DENIS0524$-$0437 are in
  Orion D (OriD) and DENIS1406$-$4151 in Upper Centaurus-Lupus (UCL), as discussed
  concerning their membership in young associations below, and we then used
  average distances of 350$\pm$3~pc for OriD (\citealt{kounkel18})
  and 130$\pm$20~pc for UCL (e.g., see \citealt{gagne18} and references therein)
  to estimate the ages and masses of these three objects.
Our estimates indicate that the masses of 21 candidates are below the substellar mass, and the objects are therefore BD candidates. The masses for 2 of them are above 75~$M_{\rm Jup}$, and they are
VLM stars.
Figure~\ref{f1} also shows that according to the BT-Settl CIFIST 2011-2015 models,
lithium is still present and detectable
in the stellar atmospheres of all 23 candidates.

For each candidate,
we also modeled a blackbody to estimate the disk mass $M_{\rm disk}$,
the disk temperature $T_{\rm disk}$, 
and the fractional IR luminosity $f = L_{\rm IR}/L_{\star}$.
The DENIS $I$-band, 2MASS $J$, $H$, and $K_{\rm S}$-band,
and WISE $W1$-band fluxes were used to
find the best-fit BT-Settl stellar atmosphere models.
The results indicate low surface gravities $\log g$ from 3.0 to 4.5
in all candidates.
Most of our candidates only show IR excesses in WISE $W3$ and $W4$-bands
where the disk emission predominates in comparison to the stellar emission. 
We therefore only used the WISE $W3$ and $W4$-band fluxes to model the disk.
Table~\ref{mass} lists the results for the 23 candidates.
The best fits of the disk blackbody modeling to the photometric data are shown in Figure~\ref{seda}.
  The SEDs of DENIS0534$-$1048, DENIS1334$-$4619, DENIS1448$-$3859, DENIS1506$-$3414,
  and DENIS1546$-$2556 fit the $W1$ and $W2$ data points poorly.
  The poor fits may indicate that there
  is both a hot inner and a cold outer disk component.
  This requires a fit with two blackbodies at different temperatures,
  as discussed in \citet{morales-gut}.
  At this point, a further detailed analysis of the disk structure of these objects is needed. 
The fractional IR luminosity $f$ of all our candidates lies above the boundary of $f \sim 0.01$
between the debris and protoplanetary disks \citep{hughes,sgro},
indicating that all the 23 candidates harbor protoplanetary disks.
Ten of these detections of protoplanetary disks are new, and
13 were previously reported in the literature (see Table~\ref{mass}).

To determine the membership of these 23 candidates in young associations,
we used the Bayesian analysis tool
BANYAN~$\Sigma$\footnote{http://www.exoplanetes.umontreal.ca/banyan/banyansigma.php}
for candidates within about 150 pc \citep{gagne18}.
We also searched VizieR and SIMBAD for previously identified members
of young associations in the literature.
There are 5 new candidate members and 18 previously reported ones of young associations.
Our identification of the 18 known members based on the membership probability values
agree with the literature, except
for DENIS0338$-$3334, DENIS0338$-$2646, DENIS0844$-$7833, and DENIS0856$-$1342. 
The results based on the BANYAN tool
imply that these 4 candidates are field objects. 
The membership probabilities for the field are 51.1\%, 99.9\%, 61.2\%, and 95.8\% for
DENIS0338$-$3334, DENIS0338$-$2646, DENIS0844$-$7833, and DENIS0856$-$1342, respectively.
In contrast to the results from the BANYAN tool,
the membership probabilities from the literature (see Table~\ref{mass}), however,
imply that DENIS0338$-$3334 \citep{cantat18} and DENIS0338$-$2646 \citep{galli21}
are members of $\chi^{1}$ Fornacis (XFor),
and DENIS0844$-$7833 \citep{cantat18}
is a member of $\eta$~Chamaeleon ($\eta$~Cha).
Our detections of lithium and strong H$\alpha$ emission in these 3 objects
(see Sec.~3.2 for further details) strongly support their youth.
We therefore adopted their membership in young associations from the literature. 
For the case of DENIS0856$-$1342, our revised membership probability of this object in
TW Hydrae (TWA) with Gaia parallax and proper
motion is low, only 2.1\%. This is consistent with the value of 4.9\% estimated
by \citet{gagne15}.
Its optical \citep{cooper24} and near-IR \citep{gagne15}
spectra have exhibited signs of a low-gravity M8 dwarf, which supports its youth.
As discussed by \citet{boucher}, DENIS0856$-$1342 might be
a member of another young association that is not yet included in BANYAN. This
results in a low membership  probability in TWA.

We used the Vizier tool for the five new candidate members, three of which lack Gaia parallaxes
(DENIS0516$-$0224, DENIS0524$-$0437, and DENIS1406$-$4151)
and two of which lie at distances greater than 150 pc
(DENIS0521$-$0802 and DENIS0534$-$1048)\footnote{https://vizier.cfa.harvard.edu/viz-bin/VizieR?-source=VI/42},
which uses the constellation boundaries given by \citet{roman}
to provide the constellation in which each candidate is located. 
We found that DENIS0516$-$0224, DENIS0524$-$0437, DENIS0521$-$0802, and DENIS0534$-$1048
are located in the region of Orion (Ori)
and DENIS1406$-$4151 lies in the region of Centaurus (Cen). 
This suggests that these five M dwarfs are new candidate members of the young associations.
Further measurements of radial velocity and/or parallax of the M dwarfs are needed to confirm their membership.
Our results are listed in Table~\ref{mass}.

The VLM star
DENIS0521$-$0802 (83 $M_{\rm Jup}$) lies at a Gaia distance of 365~pc, which
is consistent with an average distance of 350$\pm$3~pc to subregions of
the OriD cloud (see \citealt{kounkel18}).
In DENIS0534$-$1048 (74 $M_{\rm Jup}$, 1 Myr), however,
the BD candidate is located at
238~pc toward the OriD cloud, but much closer than OriD. 
This implies that DENIS0534$-$1048 might be a member of a foreground population in front of
OriD (hereafter, OriD-foreground).
The distance to DENIS0534$-$1048 is
comparable to the Hipparcos distance, $\sim$221$\pm$38~pc to the supergiant $\kappa$ Ori \citep{pillitteri}.
  DENIS0534$-$1048 has a Gaia proper motion with $\mu$RA~$= 2.511\pm0.289$~mas~yr$^{-1}$
  and $\mu$DE~$= -4.320\pm0.264$~mas~yr$^{-1}$.
  These Gaia proper motion values are comparable to those of
  $\kappa$ Ori as measured by Hipparcos with $\mu$RA~$= 1.55\pm0.67$~mas~yr$^{-1}$
  and $\mu$DE~$= -1.20\pm0.50$~mas~yr$^{-1}$.
  The small difference between the proper motion values of these two objects
  may be interpreted as a spread in velocities as seen in any association.
The location of the BD candidate is about 3.4$^{\circ}$ from $\kappa$ Ori.
This suggests that the OriD-foreground population may be associated with $\kappa$ Ori. 
Further studies are needed to clarify the membership of DENIS0534$-$1048
and to confirm the existence of the OriD-foreground population. 

\begin{figure}
   \centering
    \includegraphics[width=10cm,angle=-90]{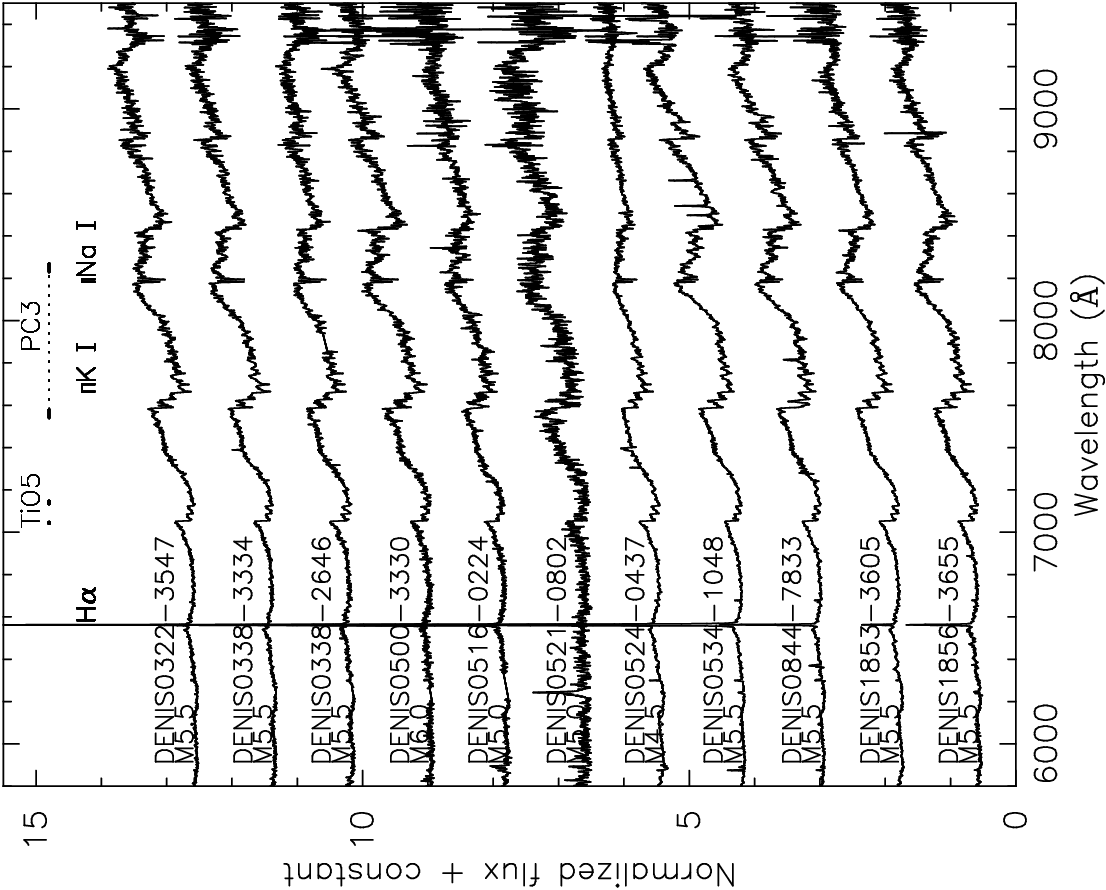}
    \caption{R3000 spectra of 11 IR excess candidates. Our spectral types
      adopted from spectral indices VOa, TiO5, and PC3 (see Table~\ref{spt}) are also shown.}
\label{f2}
\end{figure}
\begin{figure*}
   \centering
    \includegraphics[width=12cm,angle=-90]{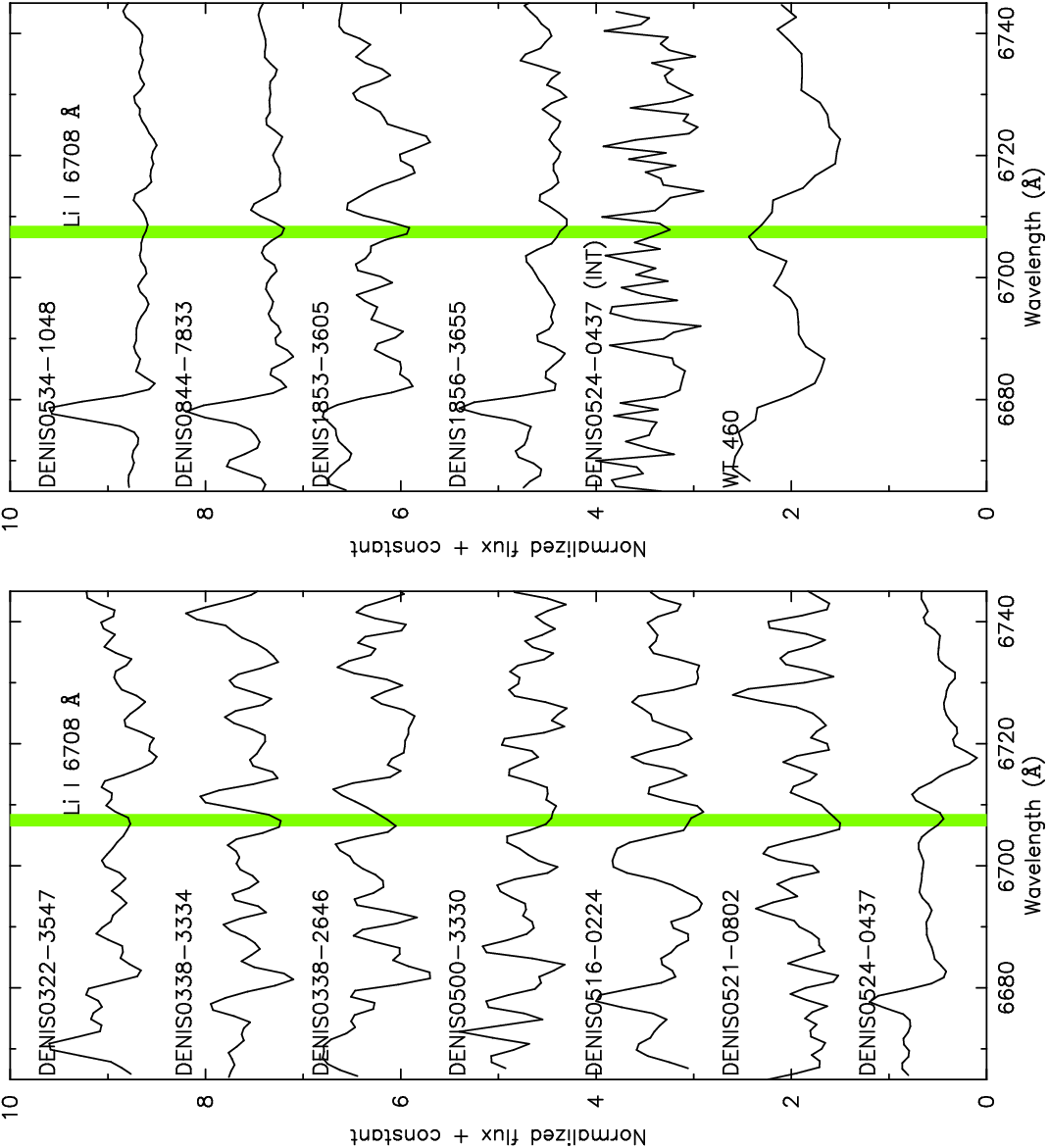}
    \caption{R3000 spectra of 11 candidates.
      The region of the Li~I line is indicated.
      The R600R INT spectrum of DENIS0524$-$0437 is shown. 
      The low-resolution spectrum of WT~460 (M5.5) without a lithium detection
      \citep{pb17} is also shown for comparison.}
\label{f3}
\end{figure*}

\subsection{Spectroscopic follow-up observations}

We observed 11 of the 23 candidates
with the Wide Field Spectrograph (WiFeS; \citealt{dopita}) on
the ANU 2.3 m telescope at Siding Spring Observatory.
We used the R3000 grating over 5300$-$9700~\AA~
with a pixel size of 1.25~\AA~ and a resolution of $R\approx3000$.
The signal-to-noise ratios per pixel of the R3000 spectra are in the range of 7$-$11.
We additionally observed 6 of the 11 candidates
with the R7000 grating
over a wavelength range of 5300$-$7050~\AA~ and a pixel size of 0.44~\AA~,
providing a spectral resolution of $R\approx7000$. 
The signal-to-noise ratios per pixel of the R7000 spectra are in the range of 7$-$12.
Tables~\ref{spt} and \ref{acc} list the observing dates of the
candidates.

We then used FIGARO \citep{shortridge} for the data reduction.
The spectra were corrected for telluric absorption.
Smooth spectrum stars were additionally observed
to remove the telluric lines (see \citealt{bessell99}). 
A NeAr arc was used for the wavelength calibration.

The R3000 spectra of the 11 M dwarfs are shown in Figure~\ref{f2}.
Their spectral types are estimated from three indices VOa, TiO5, and PC3,
as described in \citet{pb17,pb06a}.
The adopted spectral type (Table~\ref{spt}) is an average value
calculated from the three spectral indices.
The estimated spectral types of these candidates
are in the range from M4.5 to M6.0.
Their $I-J$ colors in the range of 1.91$-$2.45 also agree with 
the unreddened $I-J$ colors of $\sim$1.9$-$2.3 of young M4.5$-$M6 dwarfs
(see \citealt{luhman04b} and references therein) within the error bars.
This consistency implies that the reddening effect is not significant for these M dwarfs. 
Therefore, we did not consider the dereddening of SEDs of
the 11 M dwarfs or the 12 remaining candidates.

\begin{figure}
   \centering
    \includegraphics[width=12cm,angle=-90]{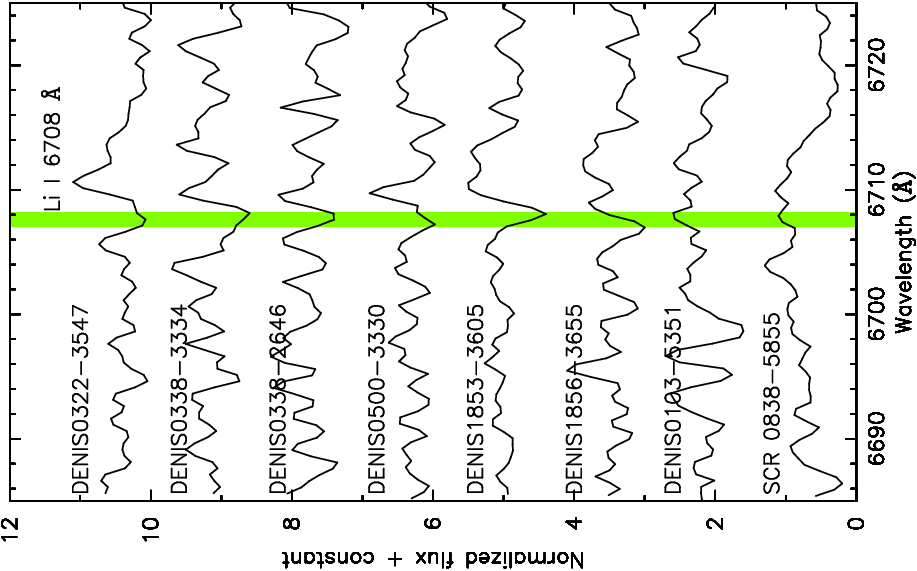}
    \caption{R7000 spectra of six candidates (see Table~\ref{acc}).
      The region of the Li~I line is indicated.
      The spectra of DENIS0103-5351 (M5.5) and SCR 0838-5855 (M6.0) without a lithium detection
      \citep{pb17} are also plotted for comparison.
}
\label{f4}
\end{figure}

\begin{figure}
   \centering
    \includegraphics[width=8cm,angle=-90]{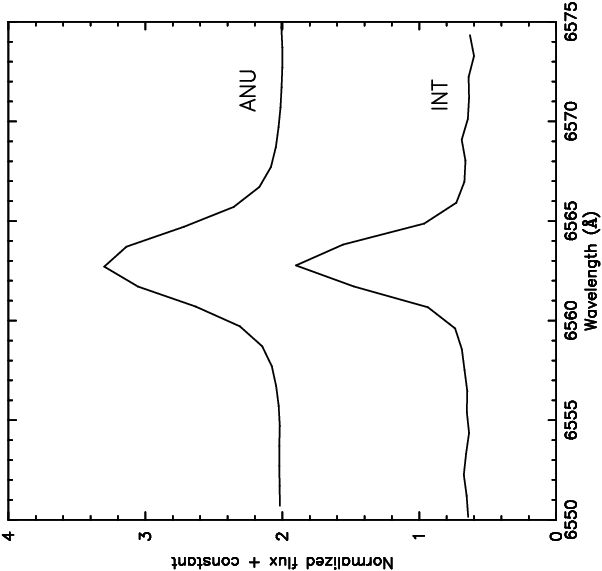}
    \caption{Spectra of DENIS0524$-$0437 observed on October 11, 2023 with ANU and December 3, 2023 with INT.      
      The change in H$\alpha$ EW measured in the ANU and INT spectra implies an accretion process in DENIS0524$-$0437
      (see Sect.~3.2 for further discussion). 
}
\label{f5}
\end{figure}

All observed 11 M dwarfs indicate the Li~I line at
  6708~\AA~ in the R3000 spectra (Fig.~\ref{f3}), implying the youth of these M dwarfs.
The lithium search for 10 of them is reported for the first time here,
and one has previously been reported in the literature (see Table~\ref{spt}).
We measured equivalent widths (EW) of the Li~I line using the IRAF task splot,
as described in the previous papers \citep{pb17,dat}.
For each spectrum, we visually determined the continuum levels and integration limits.
We estimated the EW uncertainties
by measuring EWs with different possible continuum levels and calculating
the expected uncertainty from Cayrel's formula \citet{cayrel}.
Our R7000 spectra of the six M dwarfs (Table~\ref{acc}) also show
  lithium (Fig.~\ref{f4}), confirming the lithium presence in the R3000 spectra
  of these six M dwarfs.
For DENIS0844$-$7833, the previously reported detections of lithium with EW measurements of
0.48~\AA~ \citep{song} and 0.536~\AA~ \citep{schneider}
are consistent with our value of 0.6$\pm$0.3~\AA.
In addition, we also observed DENIS0524$-$0437 on December 3, 2023
at the Isaac Newton Telescope (INT)
with the R600R grating over 4502$-$8916~\AA, corresponding to a spectral
resolution of $R\approx3400$.
The Li I EW measured in the
R600R INT spectrum is 0.4$\pm$0.2~\AA.
Our lithium detection in the R7000 spectra is consistent with the
predictions of BT-Settl models based on the Gaia parallaxes
and the $I-J$ color (see Fig.~\ref{f1}), as discussed in Sec.~3.1.

Strong H$\alpha$ emission is also detected in all M dwarfs (Fig.\ref{f2}), which
further supports their youth.
Using the same method as applied to the EW measurement of lithium,
we also measured the EWs of H$\alpha$ emission.
Table~\ref{spt} lists our measurements.
Figure~\ref{f5} shows the H$\alpha$ emission
in DENIS0524$-$0437 observed with the ANU and INT telescopes.
The H$\alpha$ EWs measured in the ANU and
INT spectra are $-33\pm4$~\AA~ and $-18\pm1$~\AA, respectively.
The change in H$\alpha$ EW implies an accretion process in DENIS0524$-$0437.

As discussed in Sec.~3.1, the best-fit BT-Settl models to the photometric data
results in low surface gravities $\log g$ in the range of 3.0$-$4.5 for
all 23 candidates and in the range of 3.5$-$4.5 for 11 M dwarfs with observed spectra
(Table~\ref{mass}).
In young M dwarfs, the Na~I doublet at 8183~\AA~ and 8199~\AA~
is an indicator of surface gravity \citep{martin96,lyo04,schlieder,pb17,dat}.
We thus measured the EWs of the Na~I doublet over the range of 8170$-$8200~\AA~
and their associated uncertainties as described by \citet{martin10} and \citet{pb17}.
Our measurements show that all 11 M dwarfs have weak Na~I (Table~\ref{spt}),
which agrees well with the modeling results of low surface gravities $\log g$.
The values of Na~I EWs of the young BD candidates in this paper are also comparable
with the Na~I EW measured in the young BD, DENIS-P~J1538317$-$103850
with a detected sporadic and intense accretion disk \citep{dat}.

\begin{figure}
   \centering
    \includegraphics[width=8cm,angle=0]{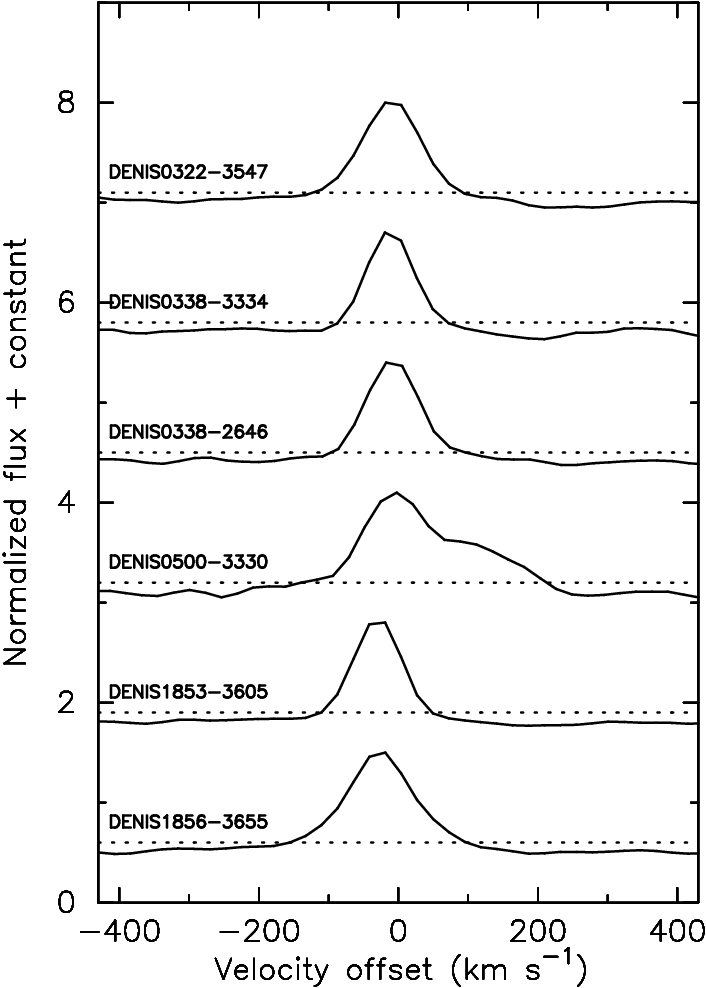}
    \caption{R7000 H$\alpha$ velocity profiles of six candidates (see Table~\ref{acc}).
      The dashed line indicates the full width at 10\% of the peak.
}
\label{f6}
\end{figure}

\begin{table*}
  \caption{Star and disk properties and membership in young associations of the 23 BD candidates and VLM stars.}
    \label{mass}
  $$
   \begin{tabular}{llllllllllll}
   \hline 
   \hline
   \noalign{\smallskip}
D* name        & Age  &  Mass        & $T_{\star}$ & $\log g$ & $T_{\rm disk}$ & $f$  & Prob.$^{\rm a}$ & Prob.$^{\rm b}$ & Membership  & Membership  & Ref.    \\
               & (Myr)&  ($M_{\rm Jup}$)&    (K)   & (cm/s$^{2}$) & (K)       &       &      &         & (this paper)& (literatue) &       \\ 
      \noalign{\smallskip}
\hline
0322$-$3547 & ~~6 &~~45$^{+ 1}_{- 1}$ & 2800 & 3.5 & 215 & 0.032 & 99.9 & 100 & XFor & XFor  & 1, 2 \\
0338$-$3334 & ~~9 &~~43$^{+ 1}_{- 1}$ & 2800 & 3.5 & 186 & 0.024 & 48.9 & 100 & XFor & XFor  &  1, 2 \\
0338$-$2646 &  10 &~~55$^{+ 5}_{- 5}$ & 2800 & 4.5 & 186 & 0.103 & ~~0  &~~98 & XFor & XFor  &  3, 2 \\
0500$-$3330 &  20 &~~45$^{+ 1}_{- 1}$ & 2700 & 3.5 & 164 & 0.063 & 57.9 &~~58 & Col  & Col  &  4, 2 \\
0516$-$0224 & ~~1 &~~73$^{+ 1}_{- 1}$ & 2700 & 3.5 & 200 & 0.079 &      &     & Ori$^{\rm c}$  &   &  2, 2 \\
0521$-$0802 & ~~1 &~~83$^{+ 5}_{- 8}$ & 2900 & 3.5 & 200 & 0.079 &      &     & Ori$^{\rm c}$  &   &  2, 2 \\
0524$-$0437 & ~~1 & 104$^{+ 2}_{- 2}$ & 2800 & 4.5 & 181 & 0.039 &      &     & Ori$^{\rm c}$  &   &  2, 2 \\
0534$-$1048 & ~~1 &~~74$^{+11}_{- 4}$ & 2900 & 3.5 & 218 & 0.183 &      &     & OriD-foreground$^{\rm c}$ &   &  2, 2 \\
0844$-$7833 & ~~1 &~~41$^{+ 1}_{- 1}$ & 2700 & 3.5 & 220 & 0.061 & 21.8 & 100$^{\rm d}$ & $\eta$ Cha & $\eta$ Cha  &  5, 6 \\
0856$-$1342 &  30 &~~37$^{+ 1}_{- 1}$ & 2600 & 3.5 & 248 & 0.068 & ~~2.1& ~~~~4.9 & TWA       & TWA  &  7, 8 \\
1334$-$4619 & ~~5 &~~29$^{+ 1}_{- 1}$ & 2600 & 3.5 & 243 & 0.087 & 53.2 &     & UCL  & Lupus III  &  9, 10 \\
1342$-$4413 & ~~6 &~~45$^{+ 1}_{- 1}$ & 2800 & 3.5 & 229 & 0.034 & 72.9 &     & UCL  & LCC        &  11, 11 \\
1344$-$4336 & ~~6 &~~46$^{+ 1}_{- 1}$ & 2800 & 3.5 & 215 & 0.087 & 87.6 &     & UCL  & Lupus III  &   9, 10 \\
1403$-$3354 & ~~4 &~~47$^{+ 1}_{- 1}$ & 2800 & 4.5 & 166 & 0.053 & 97.4 &     & UCL  & Lupus III  &   9, 10 \\
1406$-$4151 & ~~1 &~~31$^{+ 6}_{- 12}$& 2500 & 3.0 & 195 & 0.041 &      &     & Cen$^{\rm c}$  &   &   2, 2 \\
1448$-$3859 & ~~1 &~~42$^{+ 1}_{- 1}$ & 2800 & 3.5 & 239 & 0.054 & 99.7 &     & UCL  & UCL  &   9, 10 \\
1506$-$3414 & ~~6 &~~42$^{+ 1}_{- 1}$ & 2900 & 4.0 & 214 & 0.098 & 99.5 &     & UCL  & UCL  &   9, 12 \\
1546$-$2443 & ~~1 &~~52$^{+ 6}_{- 3}$ & 2800 & 3.5 & 156 & 0.059 & 98.0 &     & USco & USco &   13, 14 \\
1546$-$2556 & ~~1 &~~48$^{+ 1}_{- 1}$ & 2800 & 3.5 & 175 & 0.124 & 95.9 &     & USco & USco &   15, 15 \\
1558$-$1813 & ~~4 &~~48$^{+ 1}_{- 1}$ & 2800 & 3.5 & 206 & 0.044 & 99.4 &     & USco & USco &   13, 16 \\
1605$-$1818 & ~~1 &~~39$^{+ 1}_{- 1}$ & 2700 & 3.5 & 211 & 0.098 & 99.7 &     & USco & USco &   17, 18 \\
1853$-$3605 & ~~3 &~~72$^{+ 1}_{- 1}$ & 2800 & 4.5 & 218 & 0.051 & 99.3 &     & UCrA & UCrA &   19, 2 \\
1856$-$3655 & ~~6 &~~45$^{+ 1}_{- 1}$ & 2800 & 3.5 & 190 & 0.065 & 96.5 &     & UCrA & CrA  &   20, 21 \\
    \noalign{\smallskip}
    \hline 
   \end{tabular}
  $$
  \begin{list}{}{}
  \item[] 
    {\bf Notes.} The uncertainties for the estimated values of effective temperatures $T_{\star}$
    and surface gravities $\log g$ are 100~K and 0.5, respectively.
    $^{\rm (a)}$ The membership probability using the BANYAN~$\Sigma$ tool.
    $^{\rm (b)}$ The membership probability from the literature.
    $^{\rm (c)}$ See Sec.~3.1 for further discussion.
    $^{\rm (d)}$ The membership probability from \citet{cantat18}. \\
{\bf References.} References for membership and disk detection: (1) \citet{cantat18}; (2) this paper;
(3) \citet{galli21}; (4) \citet{sarro}; (5) \citet{song}; (6) \citet{meg}; (7) \citet{gagne15};
(8) \citet{boucher}; (9) \citet{damiani};
(10) \citet{luhman22}; (11) \citet{goldman}; (12) \citet{teixeira}; (13) \citet{lodieu13}; (14) \citet{esplin18};
(15) \citet{dawson}; (16) \citet{theissen}; (17) \citet{ardila}; (18) \citet{morales}; (19) \citet{galli20};
(20) \citet{peterson}; (21) \citet{esplin22}.
  \end{list}
\end{table*}

\begin{table*}
  \caption{H$\alpha$, Li~I 6708 \AA, and Na~I (8170-8200~\AA) EWs,
    spectral indices and spectral types estimated from the R3000 spectra for 11 young BD candidates and VLM stars.}
    \label{spt}
  $$
   \begin{tabular}{llllllllll}
   \hline 
   \hline
   \noalign{\smallskip}
D* name              & UT observing & EW H$\alpha$ & EW Li & EW Na I  & VOa  & TiO5 & PC3 &  SpT$^{\rm a}$ & Ref. \\
                  &   date       &   (\AA)            &   (\AA)     &  (\AA)    &      &     &      &               &  \\ 
      \noalign{\smallskip}
\hline
0322$-$3547 & 2023-09-06 &~~$-$16$\pm$1  & 0.6$\pm$0.3 & 4.7$\pm$0.3 & 2.07$\pm$0.18 & 0.22$\pm$0.03 & 1.34$\pm$0.10 & M5.5 & \\
0338$-$3334 & 2023-08-29 &~~$-$16$\pm$1  & 0.7$\pm$0.3 & 4.8$\pm$0.2 & 2.11$\pm$0.18 & 0.26$\pm$0.04 & 1.38$\pm$0.07 & M5.5 &  \\
0338$-$2646 & 2023-09-06 &~~~~$-$8$\pm$1 & 1.1$\pm$0.3 & 5.3$\pm$0.3 & 2.07$\pm$0.23 & 0.24$\pm$0.03 & 1.27$\pm$0.09 & M5.5 & \\
0500$-$3330 & 2023-09-29 &~~$-$60$\pm$5  & 0.3$\pm$0.4 & 4.7$\pm$0.3 & 2.18$\pm$0.28 & 0.23$\pm$0.06 & 1.43$\pm$0.12 & M6.0 & \\
0516$-$0224 & 2023-10-07 &~~$-$60$\pm$5  & 0.6$\pm$0.3 & 4.0$\pm$1.1 & 2.07$\pm$0.21 & 0.41$\pm$0.08 & 1.32$\pm$0.12 & M5.0 &  \\
0521$-$0802 & 2023-10-03 &~~$-$15$\pm$1  & 2.9$\pm$0.7 & 2.8$\pm$1.4 & 2.01$\pm$0.43 & 0.24$\pm$0.11$^{\rm b}$ & 1.30$\pm$0.34 & M5.0 & \\
0524$-$0437 & 2023-10-11 &~~$-$33$\pm$4  & 0.5$\pm$0.3 & 3.4$\pm$0.4 & 2.01$\pm$0.14 & 0.36$\pm$0.02 & 1.17$\pm$0.05 & M4.5 & \\
0534$-$1048 & 2023-10-11 & $-$246$\pm$10 & 0.2$\pm$0.3 & 2.4$\pm$1.2 & 2.14$\pm$0.20 & 0.31$\pm$0.03 & 1.41$\pm$0.10 & M5.5 & \\
0844$-$7833 & 2023-10-12 &~~$-$37$\pm$3  & 0.6$\pm$0.3 & 3.6$\pm$1.0 & 2.11$\pm$0.19 & 0.27$\pm$0.02 & 1.39$\pm$0.10 & M5.5 & 1 \\
1853$-$3605 & 2023-08-04 &~~$-$10$\pm$1  & 0.7$\pm$0.3 & 3.5$\pm$1.2 & 2.05$\pm$0.16 & 0.25$\pm$0.02 & 1.30$\pm$0.08 & M5.5 & \\
1856$-$3655 & 2023-08-24 &~~$-$23$\pm$2  & 0.7$\pm$0.3 & 3.7$\pm$0.5 & 2.10$\pm$0.21 & 0.26$\pm$0.03 & 1.35$\pm$0.09 & M5.5 &  \\
    \noalign{\smallskip}
    \hline 
   \end{tabular}
  $$
  \begin{list}{}{}
  \item[] 
{\bf Notes.} 
$^{\rm (a)}$ An uncertainty of 0.5 subtypes of the estimated spectral types.  
$^{\rm (b)}$ An unreliable value that was not used to estimate spectral type.\\ 
{\bf References.} References for lithium detection: (1) \citet{song}.
  \end{list}
\end{table*}

\begin{table*}
  \caption{H$\alpha$ and Li~I 6708 \AA~ EWs, velocity width $v_{10}$, and accretion rate $\dot{M}$
    estimated from the R7000 spectra for six young BD candidates.}
    \label{acc}
  $$
   \begin{tabular}{llllll}
   \hline 
   \hline
   \noalign{\smallskip}
D* name              & UT observing & EW H$\alpha$ & EW Li & $v_{\rm 10}$(H$\alpha$) & $\log$($\dot{M}$)    \\
                  &   date       &   (\AA)            &   (\AA)     & (km s$^{-1}$)           & ($M_{\odot}$yr$^{-1}$)         \\ 
      \noalign{\smallskip}
\hline
0322$-$3547 & 2023-07-14 &~~$-$8.2$\pm$0.2 & 0.6$\pm$0.1 & 215$\pm$13 & $-$10.8$\pm$0.1  \\
0338$-$3334 & 2023-07-14 & $-$10.1$\pm$0.5 & 1.3$\pm$0.2 & 158$\pm$2  &   \\
0338$-$2646 & 2023-07-12 & $-$10.3$\pm$0.5 & 0.5$\pm$0.1 & 194$\pm$19$^{\rm a}$ & $-$11.0$\pm$0.2  \\
0500$-$3330 & 2023-07-14 & $-$30.0$\pm$2.0 & 0.5$\pm$0.1 & 352$\pm$11 &~~$-$9.5$\pm$0.1  \\
1853$-$3605 & 2023-07-14 &~~$-$8.7$\pm$0.2 & 0.8$\pm$0.1 & 159$\pm$8  &   \\
1856$-$3655 & 2023-07-14 & $-$16.4$\pm$0.4 & 0.8$\pm$0.1 & 254$\pm$7  & $-$10.4$\pm$0.1  \\
    \noalign{\smallskip}
    \hline 
   \end{tabular}
  $$
  \begin{list}{}{}
  \item[]
    {\bf Notes.} 
$^{\rm (a)}$ A potential accretor.  
  \end{list}
\end{table*}

The spectra of all 11 young M dwarfs show strong H$\alpha$ emission.
They are thus potential accretors.
The H$\alpha$ emission profile 
was used to distinguish between accretors and nonaccretors
\citep{martin98,barrado03,jay,white}.
For spectra with a moderate signal-to-noise ratio and a low resolution of young stars and BDs, 
\citet{barrado03} showed that the EW values of H$\alpha$ emission
can be used to detect accretion.  
For our low-resolution R3000 spectra of the 11 candidates,
we therefore used the criterion of H$\alpha$ EW versus spectral type
as given in \citet{barrado03} (see their Table 1)
to identify accretors.
Our observational results are listed in Table~\ref{spt}, and
five of them are accretors: DENIS0500$-$3330 (M6.0),
DENIS0516$-$0224 (M5.0), DENIS0524$-$0437 (M4.5), DENIS0534$-$1048 (M5.5),
and DENIS0844$-$7833 (M5.5).
DENIS1856$-$3655 with a spectral type of M5.5 and an H$\alpha$ EW
of $-$23$\pm$2~\AA~ is a potential accretor because
its H$\alpha$ EW value is close to the limit value of
24.1~\AA~(Table 1, \citealt{barrado03})
for an M6.
Since medium- and high-resolution spectroscopic
observations of the H$\alpha$ emission profile 
can be used to distinguish between accretors and nonaccretors
and also to measure the accretion rate \citep{jay,white},
we also used our R7000 spectra to identify accretors
and measure the accretion rate.
We measured the velocity width $v_{10}$ of the H$\alpha$ emission profile 
at 10\% of the peak flux and then used 
a cutoff of $v_{10} \sim 200$~km~s$^{-1}$,
as proposed by \citet{jay},
to separate accreting and nonaccreting M dwarfs.
Table~\ref{acc} lists our measurements of $v_{10}$.
Three young BD candidates, DENIS0322$-$3547, DENIS0500$-$3330, and DENIS1856$-$3655,
are accretors. DENIS0338$-$2646 with $v_{10}=194\pm19$~km~s$^{-1}$ close to the cutoff
is a potential accretor.
By combining the low- and medium-resolution spectra, we
identified seven accretors and one potential accretor.
The R3000 spectra with a resolution of $\sim$3000
are probably not high enough to properly measure the accretion rate, especially
in mid- and late-M dwarfs, because TiO bands at 6569~\AA~ \citep{kirk91} significantly
contaminate the H$\alpha$ emission.
We therefore only used the R7000 spectra to measure the accretion rate in accreting
BD candidates. Our measurements of the accretion rate are listed in Table~\ref{acc}.

Figure~\ref{f6} shows the R7000 H$\alpha$ velocity profiles of the six young BD candidates.
The member of Columba,
DENIS0500$-$3330, also shows an inverse-P Cygni H$\alpha$ profile,
as observed in young BDs \citep{mu03}, indicating 
magnetospheric accretion \citep{edwards}.
The accretion rate in this young BD is 10$^{-9.5}$$M_{\odot}$yr$^{-1}$, which is
comparable to the young BD ISO-Oph 102 ($\sim$1 Myr) with an accretion
rate of 10$^{-9.0}$$M_{\odot}$yr$^{-1}$ \citep{natta04}
and lies in the range of accretion rates of 10$^{-10.5}$$-$10$^{-7.9}$$M_{\odot}$yr$^{-1}$
in the 1 Myr old BD DENIS-P~J1538317$-$103850 \citep{dat}. 
Our estimated age of DENIS0500$-$3330 is 
20~Myr, indicating that the accretion and the associated outflow in
the young BD may be significantly longer-lived than in low-mass stars.
The long-lived outflow process will sweep out a significant amount of
gas from the protoplanetary disks and will thus favor
the formation of rocky planets around VLM objects (e.g., \citealt{pb08b}).

\section{Conclusion}

We presented the detection of 21 young BD candidates
(29~$M_{\rm Jup}$$-$74~$M_{\rm Jup}$) and two young VLM stars (83~$M_{\rm Jup}$, 104~$M_{\rm Jup}$)
with protoplanetary disks.
Ten of these disks are new: 3 in XFor, 1 in Col, 4 in Ori,
1 in Cen, and 1 in UCrA.
DENIS0534$-$1048 might be the first BD candidate member
of a foreground population in front of the OriD cloud
associated with $\kappa$ Ori.
Further observations are required to clarify the membership of DENIS0534$-$1048
and confirm the existence of the OriD-foreground population. 
For 3 candidates without Gaia parallaxes,
DENIS0516$-$0224 and DENIS0524$-$0437 in Ori, and
DENIS1406$-$4151 in Cen, trigonometric parallaxes are also
needed to confirm their membership.
Our spectroscopic observations revealed
lithium in six young BD candidates,
which supports their youth, as predicted by the theoretical models.
Based on our low- and medium-resolution spectra, we
identified seven accretors:
DENIS0322$-$3547 (M5.5, XFor)
DENIS0500$-$3330 (M6.0, Col),
DENIS0516$-$0224 (M5.0, Ori),
DENIS0524$-$0437 (M4.5, Ori),
DENIS0534$-$1048 (M5.5, OriD-foreground),
DENIS0844$-$7833 (M5.5, $\eta$ Cha),
and DENIS1856$-$3655 (M5.5, UCrA)
and one potential accretor: DENIS0338$-$2646 (M5.5, XFor). 
The long-lived accretion and probably the associated long-lived
outflow in the 20 Myr old BD candidate, DENIS0500$-$3330 provide additional evidence that favors the formation of rocky planets around
VLM stars and BDs.

\begin{acknowledgements}
We would like to thank the referee for useful comments.
This research is funded by Vietnam National Foundation for Science
and Technology Development (NAFOSTED) under grant number 103.99-2020.63.
ELM is supported by the European Research Council Advanced grant SUBSTELLAR, project number 101054354.
This publication makes use of data products from the Wide-field Infrared Survey Explorer,
which is a joint project of the University of California, Los Angeles,
and the Jet Propulsion Laboratory/California Institute of Technology,
funded by the National Aeronautics and Space Administration.
This work has made use of data from the European Space Agency (ESA) mission
Gaia (\url{https://www.cosmos.esa.int/gaia}), processed by the Gaia
Data Processing and Analysis Consortium (DPAC,
\url{https://www.cosmos.esa.int/web/gaia/dpac/consortium}). Funding for the DPAC
has been provided by national institutions, in particular the institutions
participating in the Gaia Multilateral Agreement.
The DENIS project has been partly funded by the SCIENCE and the HCM plans of
the European Commission under grants CT920791 and CT940627.
It is supported by INSU, MEN and CNRS in France, by the State of Baden-W\"urttemberg 
in Germany, by DGICYT in Spain, by CNR in Italy, by FFwFBWF in Austria, by FAPESP in Brazil,
by OTKA grants F-4239 and F-013990 in Hungary, and by the ESO C\&EE grant A-04-046.
Jean Claude Renault from IAP was the Project manager.  Observations were  
carried out thanks to the contribution of numerous students and young 
scientists from all involved institutes, under the supervision of  P. Fouqu\'e,  
survey astronomer resident in Chile. 
This publication makes use of data products from the Two Micron All Sky Survey, which is a joint project of
the University of Massachusetts and the Infrared Processing and Analysis Center/California Institute of Technology,
funded by the National Aeronautics and Space Administration and the National Science Foundation.
This research has made use of the VizieR catalogue access tool, CDS,
Strasbourg, France. The original description of the VizieR service was
published in A\&AS 143, 23.
This research has made use of the SIMBAD database,
operated at CDS, Strasbourg, France.
\end{acknowledgements}

\bibliographystyle{aa}
\bibliography{ngoc}

\begin{appendix}
\onecolumn
  \section{SED of late-M dwarfs}
\begin{figure*}[h!]
  \centering
     \includegraphics[width=17.0cm,angle=0]{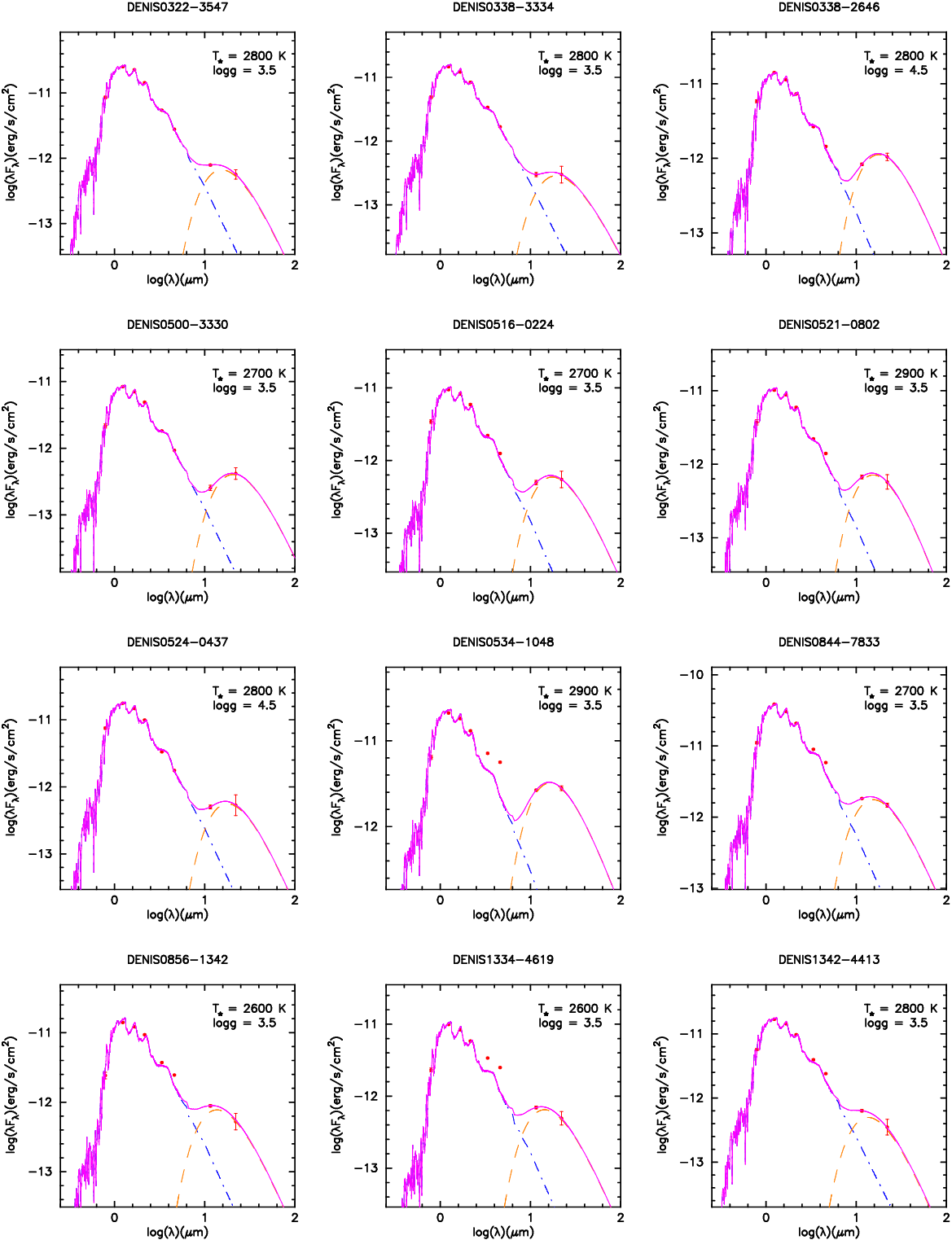}       
     \caption{SEDs for all 23 candidates with detected IR excesses.
       The blue dash-dotted line shows the BT-Settl model atmosphere.
       The disk blackbody model is indicated by the orange dashed line.
       Our best fit to the photometric data is shown by the magenta solid line.
       Effective temperature $T_{\star}$ and surface gravity $\log g$ estimated
       from the BT-Settl model are also shown.
}
\label{seda}
\end{figure*}

\begin{figure*}[h!]
  \setcounter{figure}{0}
   \centering
     \includegraphics[width=17.0cm,angle=0]{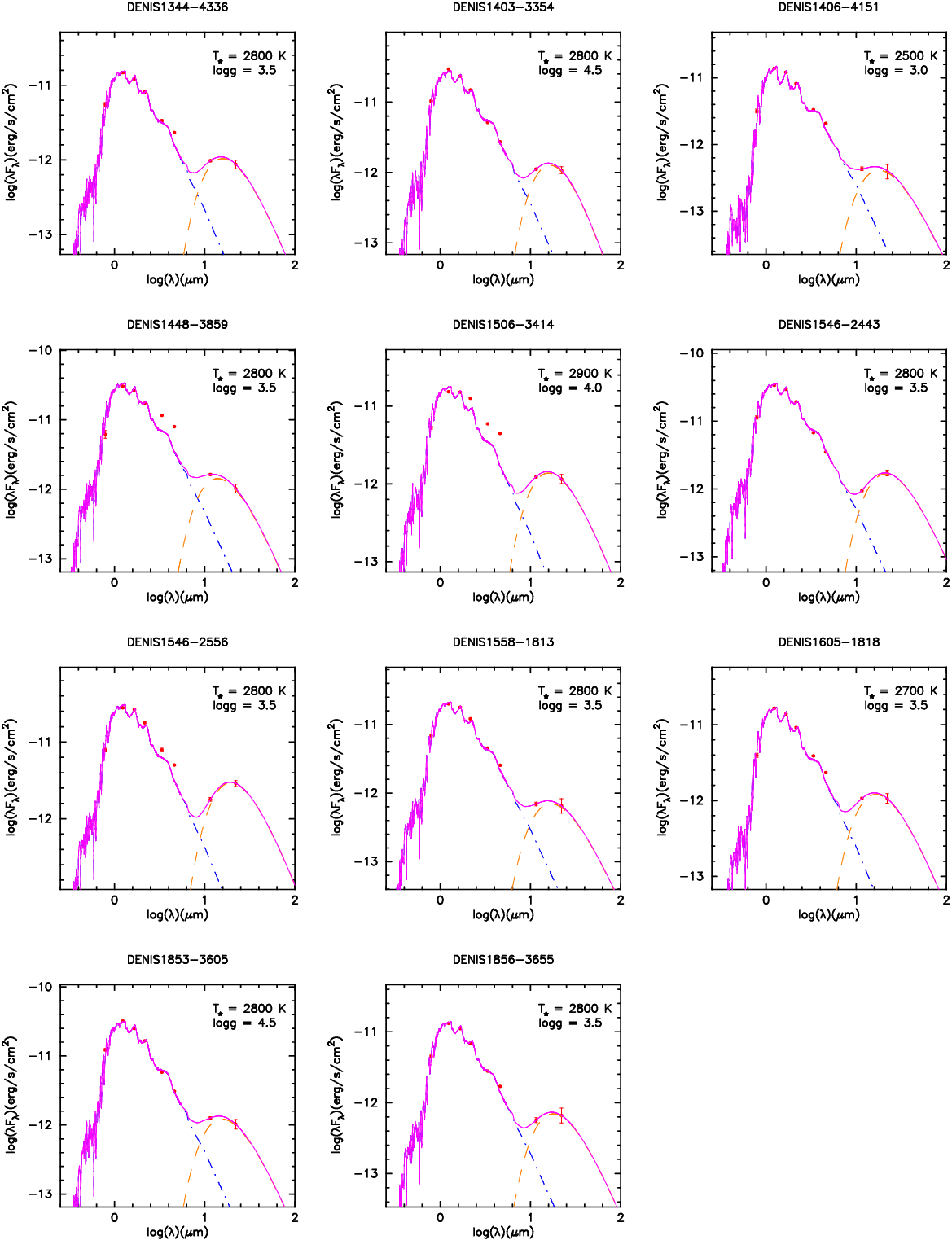}       
     \caption{continued.
}
\label{sedb}
\end{figure*}
\end{appendix}
\end{document}